\documentclass{ifacconf}

\usepackage{amsmath}
\usepackage{amsfonts}
\usepackage{dsfont}
\usepackage{graphicx}      
\usepackage{natbib}        
\usepackage[utf8]{inputenc}
\inputencoding{utf8}
\usepackage{color}
\usepackage{subfigure}
\begin{document}

\begin{frontmatter}

\title{Identification of the nonlinear steering dynamics of an autonomous vehicle\thanksref{footnoteinfo}} 

\thanks[footnoteinfo]{The research presented in this paper was carried out as part of the “Dynamics and Control of Autonomous Vehicles meeting the Synergy Demands of Automated Transport Systems (EFOP-3.6.2-16-2017-00016)” project in the framework of the New Széchenyi Plan. The research was also supported by the Ministry of Innovation and Technology NRDI Office within the framework of the Autonomous Systems National Laboratory Program}

\author[First,Second]{G. R\"od\"onyi}, 
\author[Third]{G. I. Beintema},
\author[Second,Third]{R. Tóth},
\author[Third]{M. Schoukens},
\author[First]{D. Pup} ,
\author[First,Second]{Á. Kisari},
\author[Second]{Zs. Vígh},
\author[First]{P. Kőr\"os} ,
\author[First,Second]{A. Soumelidis},
\author[Second]{J. Bokor}

\address[First]{Széchenyi István University, Research Center of Vehicle Industry \newline
(SZE-JKK) H-9026 Egyetem tér 1. Győr, Hungary.}
\address[Second]{Systems and Control Laboratory, Institute for Computer Science and Control (SZTAKI) (e-mail: soumelidis@sztaki.hu).}
\address[Third]{Control Systems, Eindhoven University of Technology, Eindhoven, The Netherlands (e-mail: r.toth@tue.nl).}

\begin{abstract}                
Automated driving applications require accurate vehicle specific models to precisely predict and control the motion dynamics. However, modern vehicles have a wide array of digital and mechatronic components that are difficult to model, manufactures do not disclose all details required for modelling and even existing models of subcomponents require coefficient estimation to match the specific characteristics of each vehicle and their change over time.
Hence, it is attractive to use data-driven modelling to capture the relevant vehicle dynamics and synthesise model-based control solutions. In this paper, we address identification of the steering system of an autonomous car based on measured data. We show that the underlying dynamics are highly nonlinear and challenging to be captured, necessitating the use of data-driven methods that fuse the approximation capabilities of learning and the efficiency of dynamic system identification. We demonstrate that such a neural network based subspace-encoder method can successfully capture the underlying dynamics while other methods fall short to provide reliable results. \vspace{-3mm}
\end{abstract}


\end{frontmatter}

\section{Introduction} \vspace{-2mm}

Promising benefits of using autonomous road vehicles, such as higher level of safety, energy efficiency, reduced emission and congestion; travel time saving (see \cite{anderson2014autonomous,Trommer16,KOLAROVA2019}), motivated technological innovations and research for decades. Transferring control and responsibility from human driver to computers demands increased level of reliability and safety in automotive control systems. Advanced control system design is model-based. The vehicle however is a complex, high dimensional, time-varying and hybrid nonlinear system with coupled components and uncertain, varying parameters, operating in a yet more and more complex and changing environment. Thus, modelling and control design are challenging tasks. To support model based control design, the dominant modelling paradigm is to build first-principles based models using physical equations \citep{Berntorp:Model, KienckeNielsen2000}. Physical parameters of such models can often be estimated on the fly and utilised in an adaptive control setting \citep{Singh:Estimation}.

Alternative model building techniques that rely more on measured data are applied when the describing equations are too complex for control design, the uncertainty in some components of the system are too large, or the conditions vary in time. The vehicles have digital and mechatronic components that are difficult to model and often the manufacturers do not disclose all details. Hence identifying a part or the whole of the behaviour of interest by means of low-order model structures is a reasonable approach. For example, \cite{RosoliaRacing} extended the known part of a discrete-time state-space model by and additive polynomial model whose coefficients were estimated by least square methods. In a similar approach an unknown model component was characterised by a \emph{Gaussian process} (GP) model, \cite{HewingZeilinger}. 

The goal of this paper is to identify a control oriented model for the lateral dynamics of a Nissan Leaf that was modified to become a platform for autonomous driving research. To automatize steering of the vehicle, the built-in servo system is utilized. In normal operation the servo system receives a voltage signal proportional to the measured torque applied by the driver. With a minimal cost hardware modification, this connection is augmented: the autonomous navigation controller running on an external computer may produce an additional voltage input to the servo system generating torque to autonomously steer the system. This concept worked well with a base-line controller as demonstrated in \cite{IFAC2020Nissan}, but a more accurate model-based controller is required to increase performance and reduce the strain to the servo. One challenge in this modeling problem is that the behavior of the servo system including its control software and mechatronic components is unknown. The other challenge lays in the nonlinear/time varying characteristics of the pneumatic trail. It causes negative self-aligning torque at large steering angles and low speed, and this makes the steering mechanics of the vehicle sensitive to disturbances.

The contributions of the paper are the following. We analyze the dynamic aspects of the given system, detail the experimental scenarios and we compare identification methods in capturing the dynamics of the system. Our analysis shows that the system is highly nonlinear and in fact challenging for system identification. By comparing nonlinear identification methods, we demonstrate that a recently introduced neural network based subspace-encoder method, which fuses the approximation capabilities of learning methods and the efficiency of subspace identification, can successfully capture the underlying dynamics while other methods fail to provide reliable results.

\vspace{-1mm}
\section{System Description}		\label{sec:sys}
\vspace{-2mm}
\subsection{Overview}
\vspace{-2mm}
The lateral dynamics of the Nissan Leaf-based autonomous vehicle are controlled by using the built-in steering servo assist unit which originally receives the driver's steering wheel torque as input and generates additional torque on the steering system. With the least intervention in the hardware, the wired connection from steering-wheel torque sensor to servo system is augmented by the possibility of superimposing an additional artificial torque signal generated by an on-board computer. In autonomous vehicle experiments, the vehicle is running with released hand-wheel, thus the sole input to the steering actuator is the requested torque signal $u_\mathrm{s}$. While this concept allows to assist or to fully take over steering from the driver, it also includes the sensor, the connected digital hardware and the overall servo dynamics between the steering system and the actuation input, which are difficult to model as (a) there is no reliable documentation available from the manufacturer and (b) it is significantly vehicle specific.

\begin{figure}
\begin{center}\vspace{-2mm}
\includegraphics[width=0.8\columnwidth]{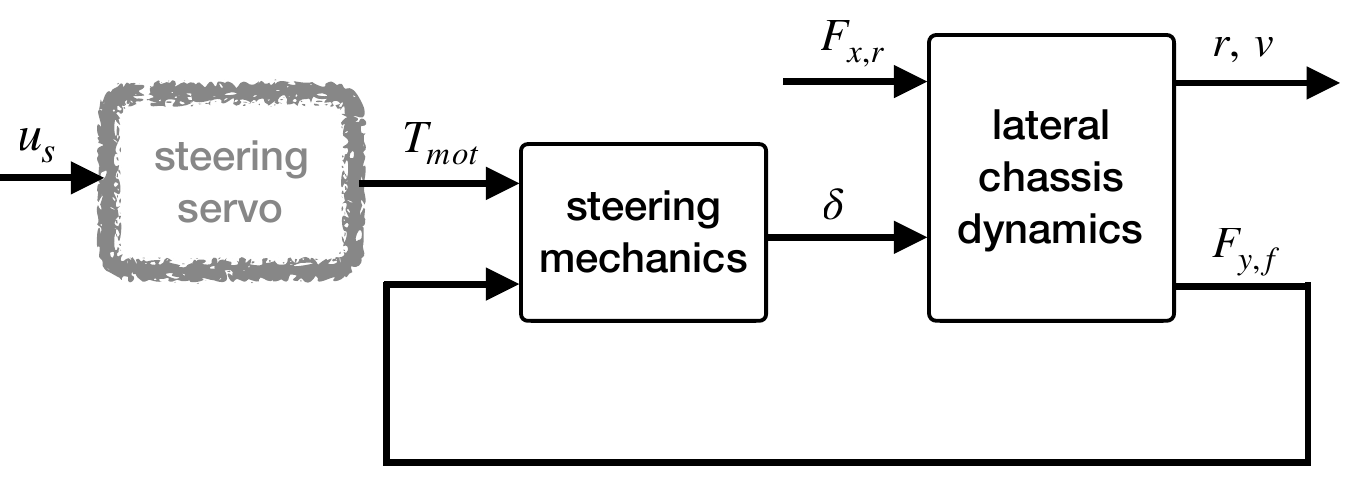}   \vspace{-3mm} 
\caption{Main components of the system. Measured data are available for $u_\mathrm{s}, \delta, r$ and $v$. } 
\label{fig:overalscheme}
\end{center}
\end{figure}

The main components of the overall steering system including the actuation are depicted in Fig. \ref{fig:overalscheme}. The lateral chassis dynamics and the steering system mechanics are generally well understood and first principles based models of these components with various complexity are available in the literature, see \cite{Berntorp6dof,KienckeNielsen2000}, which can characterize the response of these components under normal driving conditions, i.e., with moderate lateral acceleration and speed. However, often these models only capture the dominant part of the overall behavior and they require the estimation of several coefficients based on dedicated experiments. The third component, the steering assist system (servo) together with its control algorithm; however, is completely unknown and its exact design is not disclosed by the manufacturer. Hence, in overall, it becomes important to derive an accurate data-driven model of the steering system which is (i) capable to capture the unknown dynamics of the steering assist system, (ii) beyond the dominant lateral chassis dynamics and the steering system mechanics, describes accurately the specific mechanical behavior of the car, (iii) avoids the need of dedicated experiments to estimate the coefficients of first principle models and makes possible online refinement and automated periodic maintenance of the model.
In the following subsections the components are detailed and challenges of the modeling are highlighted.

\vspace{-2mm}
\subsection{Nissan Leaf Autonomous Car Prototype}
\begin{figure}[t]
	\centering 
	\includegraphics[width=0.8\columnwidth]{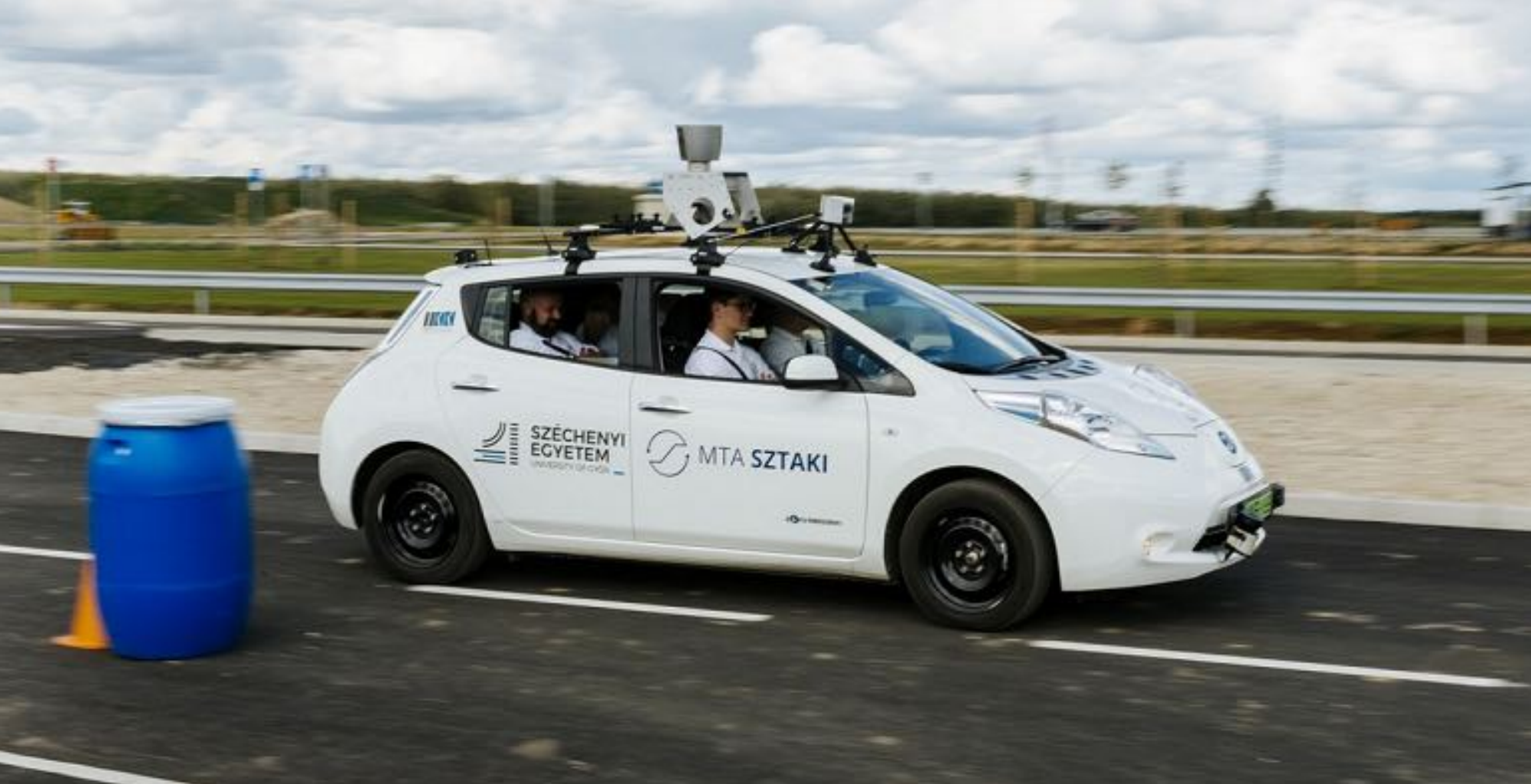} \vspace{-1mm}
	\caption{The Nissan Leaf based autonomous car.}
	\label{fig:leaf}
\end{figure}

\begin{figure}[t]
	\centering
	\includegraphics[width=0.95\columnwidth]{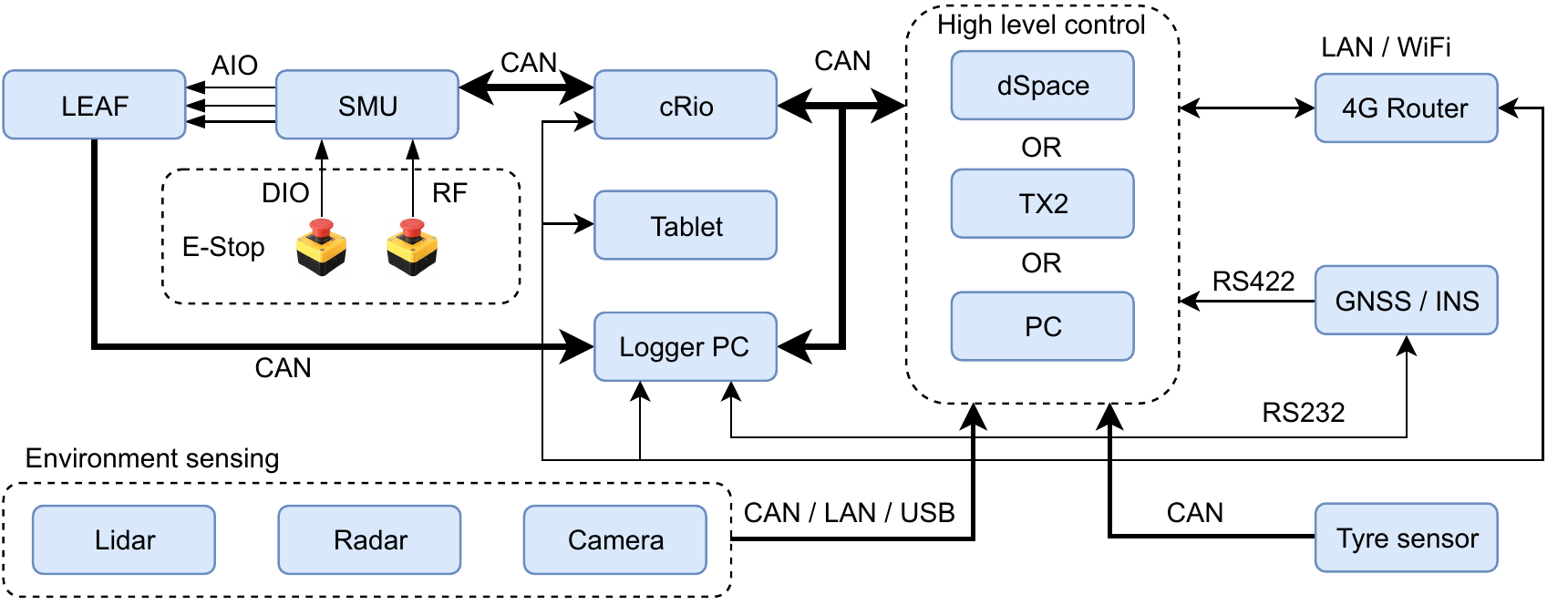} \vspace{-2mm}
	\caption{Component architecture of the autonomous car.}
	\label{fig:diagram}
\end{figure}

\vspace{-2mm}
To support research on the field of autonomous and cooperative driving a prototype vehicle platform was developed by SZTAKI and SZE-JKK. The electric drive Nissan Leaf shown in Fig- \ref{fig:leaf} was equipped with interconnected sensors and computers as illustrated in Fig. \ref{fig:diagram}. Any steering command initiated from either the NI cRIO9039 device or dSpace MicroAutoBox II device goes through the 
\emph{safety management unit} (SMU) that disables this command in case of any  steering action by the driver. Measured signals from the sensors installed on the vehicle are available on \emph{controller area network} (CAN) buses for data logging and control. The experiments are conducted on the test track ZalaZone (https://zalazone.hu).

\vspace{-2mm}
\subsection{Lateral Chassis Dynamics}
 \vspace{-2mm}
\begin{figure}
\begin{center}
\includegraphics[width=0.8\columnwidth]{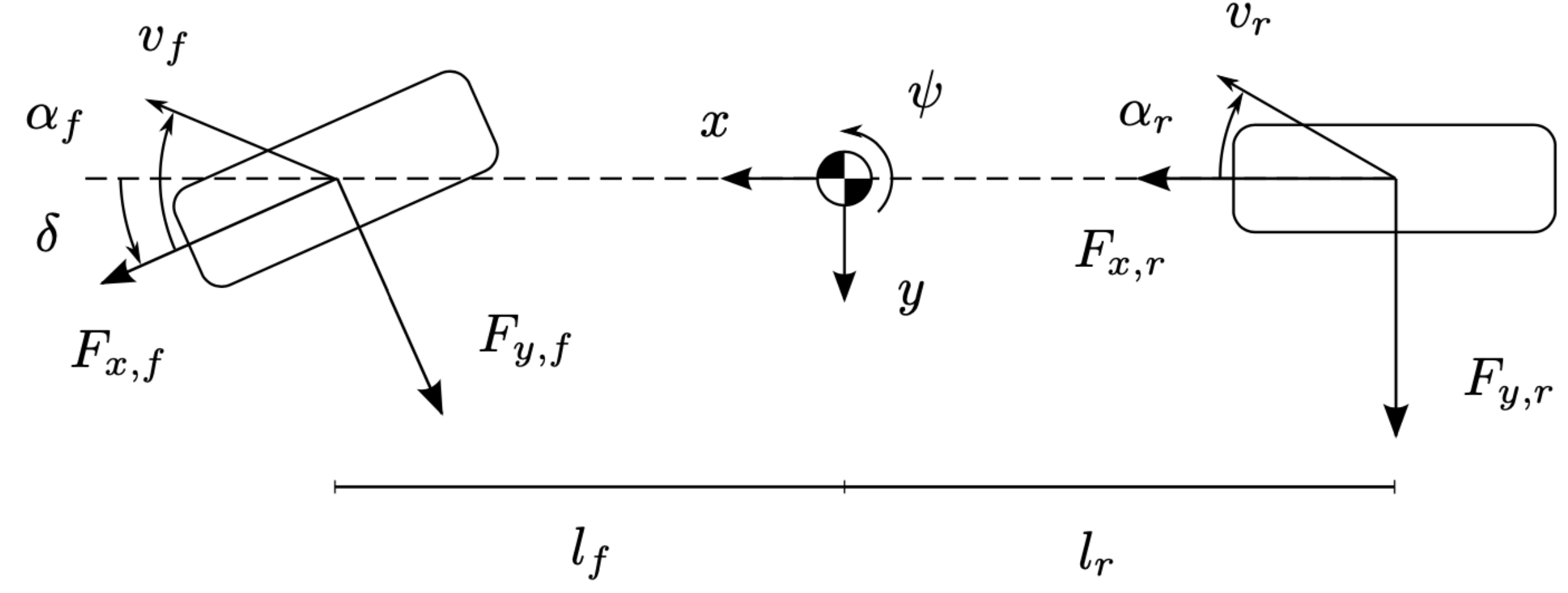}  \vspace{-2mm}   
\caption{Single-track model in \cite{Berntorp:Model}.} 
\label{fig:ST}
\end{center}
\end{figure}

The single-track chassis model describes the dominant characteristics of the translational and yaw dynamics of the vehicle at normal driving conditions \cite{Berntorp:Model}. By this simplified model, the right and left wheels are lumped together on each axle, hence roll, pitch and heave motions, and thereby load transfer, are neglected. Based on Newton's second law the state equations are 
\vspace{-1mm}
\begin{subequations}
\begin{eqnarray}
	\dot v_x(t) &\!=\!& \frac{1}{m}\bigl(F_{x,r}(t)\!-\!F_{y,f}(t)\sin(\delta(t))\!+\!m v_y(t)r(t)\bigr),			\label{eq:vx}\\
	\dot v_y(t) &\!=\!& \frac{1}{m}\bigl(F_{y,r}(t)\!+\!F_{y,f}(t)\cos(\delta(t))\!-\!m v_x(t)r(t)\bigr), 		\label{eq:vy}\\
	\dot r(t)  &\!=\!& \frac{1}{I_z}\bigl(F_{y,f}(t)l_f\cos(\delta(t))-F_{y,r}(t)l_r\bigr), 				\label{eq:r}
\end{eqnarray}
\end{subequations}
where $v_x, v_y, r$ denote the velocity components and yaw-rate in the coordinate frame of the vehicle, $m$ and $I_z$ denote mass and inertia, respectively, $l_r$ and $l_f$ denote geometric parameters according to Fig. \ref{fig:ST}. In the low wheel-slip range of moderate driving conditions, the lateral wheel forces at the rear and front can be descried as \vspace{-3.5mm} 
\begin{subequations}
\begin{eqnarray}
	F_{y,r}(t) &=& c_r\alpha_r(t), \\
	F_{y,f}(t) &=& c_f\alpha_f(t), 
\end{eqnarray}
\end{subequations}which 
depend linearly on the wheel slip angles \vspace{-2mm} 
\begin{subequations}
\begin{eqnarray}
	\alpha_f(t) &=& \delta(t) - \tan^{-1} \left(\frac{v_y(t)+v_r(t)l_f}{v_x(t)}\right), \\
	\alpha_r(t) &=& - \tan^{-1} \left(\frac{v_y(t)-v_r(t)l_r}{v_x(t)}\right).
\end{eqnarray}
\end{subequations}
In the above equations, $c_f$ and $c_r$ correspond to the so called cornering stiffness parameters. The inputs to this chassis model are the steering angle $\delta$ and the driving force $F_{x,r}$ at the rear axle. The absolute velocity, 
\begin{equation}
    v(t) = \sqrt{ v_x^2(t) + v_y^2(t) },
\end{equation}
and the yaw-rate, $r$, can be accurately measured by high precision GNSS-INS system with  a KVH GEO FOG 3D-Dual sensor. For the steering angle, $\delta$, the on-board sensor is used. For the actual car, the physical parameters $I_z$, $c_f$, $c_r$ and driving force input $F_{x,r}(t)$ are unknown.

\vspace{-2mm}
\subsection{Steering Mechanics}
\vspace{-2mm}

\begin{figure}
\begin{center}
\includegraphics[width=0.8\columnwidth]{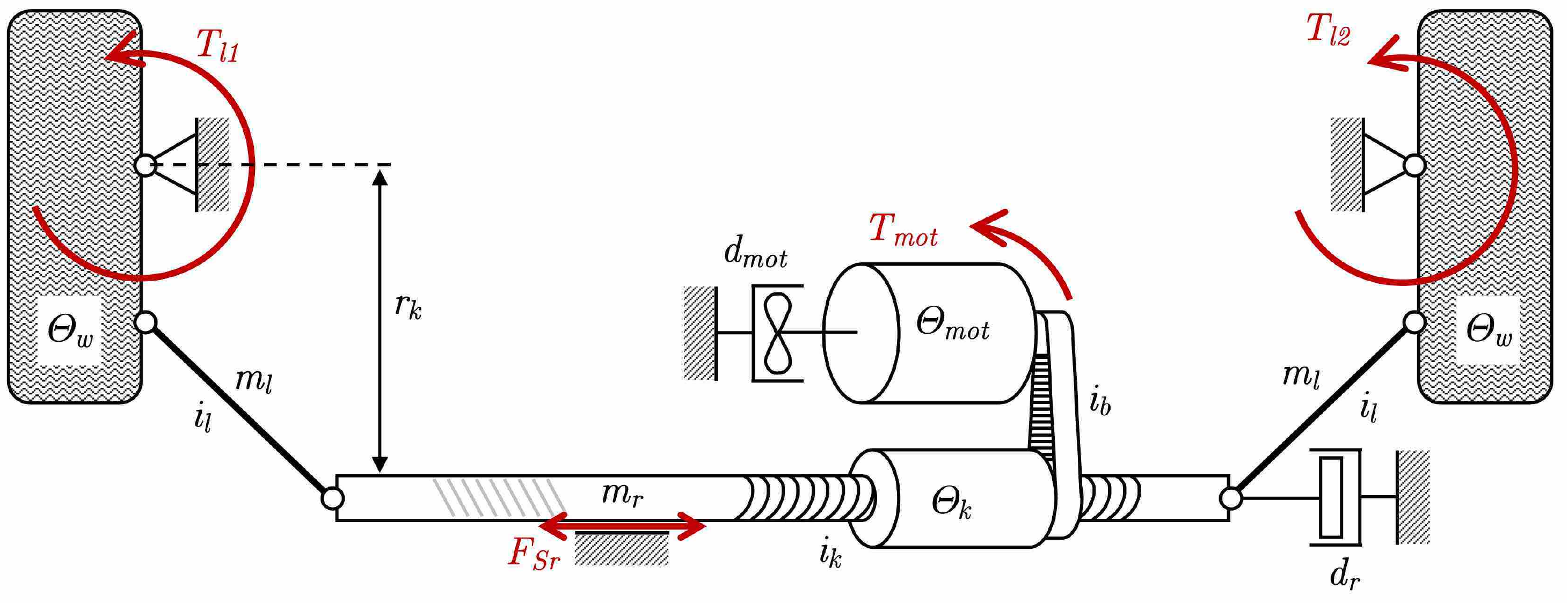}    \vspace{-2mm} 
\caption{Schematics of the steering system.}
\label{fig:SteeringSystemScheme} 
\end{center}
\end{figure}

The schematic architecture of the steering system is shown in Fig. \ref{fig:SteeringSystemScheme}. The motion of the front wheels and the linkage system are described by
\begin{equation}
	\ddot \delta(t)\Theta_\delta = -\dot\delta d_\delta + T_\mathrm{mot} + T_l - \mathrm{sign}(\delta)F_{Sr}/i_l 		\label{eq:steerdyn}
\end{equation}
where $\delta$ is the effective steering angle and $\Theta_\delta$ and $d_\delta$ correspond to the aggregated inertia and damping. The inputs influencing the steering angle are the torque $T_\mathrm{mot}$ provided by the electric servo motors, the friction force $F_{Sr}$ and the torque $T_l = T_{l,1}+T_{l,2}$ due to tire-road contact. The latter can be modeled as 
\begin{equation}
	T_l = n_{sa}F_{y,f} + T_B
\end{equation}
where $n_{sa}F_{y,f}$ is the self-aligning torque and $T_B$ is the low-speed steering friction torque between tire and pavement. For more details and analytic/empirical expressions for these torques, see \cite{StudyonLowSpeed, Analysisofvehiclestaticsteeringtorque}. 

Unfortunately, none of the terms and parameters in \eqref{eq:steerdyn} are known. The pneumatic trail $n_{sa}$, the force arm of the self-aligning torque, may vary with the steering angle and may even be negative at sharp cornering. As the trail decreases, the steering system goes toward the border of its stability region. A negative trail may imply the situation where the dominating lateral force $F_{y,f}$ turns the wheels toward their limit position (second stability region). Near the borders of the different stability regions the system is very sensitive to disturbing effects of the road and the flexible tires. This phenomenon can be observed in low speed experiments and represents a significant challenge for identification.

\vspace{-2mm}
\subsection{Modeling Approach}
\vspace{-2mm}
Looking at Fig. \ref{fig:overalscheme} and in terms of the above discussion, it can be concluded that the unknown and uncertain components in the overall system are required to be modeled. Furthermore, critical components such as the servo system and its software behavior and the pneumatic trail dynamics are completely unknown without any reliable first-principles based structure to estimate them. For this reason, we consider the system as a whole, and try to identify it in terms of nonlinear \emph{black-box} model structures\footnote{Note that considered GP and ANN model structures can be directly used in control via a wide range of model predictive (GP-MPC and ANN-MPC) solutions, e.g., see \cite{HewingZeilinger}.} presented in Section \ref{sec:ident}. The  overall system has the yaw-rate $r$ as the output while the control signal  $u_\mathrm{s}$ and the vehicle speed $v$, which represents the effect of the longitudinal dynamics on the lateral one, are considered as inputs.

\vspace{-2mm}
\section{Experimental Campaign}			\label{sec:mes}
\vspace{-2mm}
\begin{figure}
\begin{center}
\includegraphics[width=2.5cm,angle=-90]{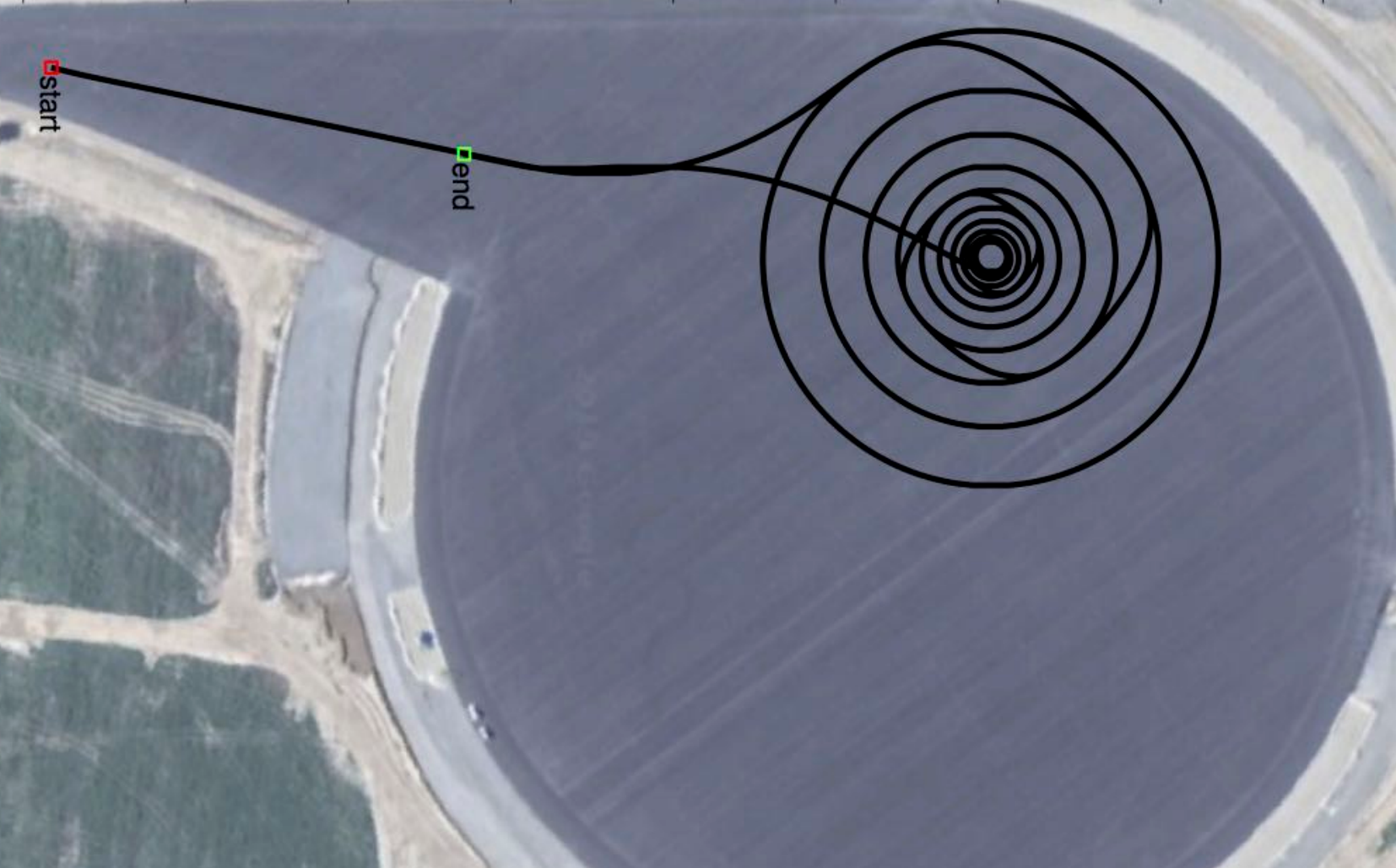}    
\caption{GPS path of a typical experiment for identification} 
\label{fig:ZalaZoneGPSident}
\end{center}
\end{figure}

Experiment design is based on three arguments: 1.) Physical insight and driving experience show that lateral dynamics vary over regions of operation. The simplified lateral chassis dynamics \eqref{eq:vy}-\eqref{eq:r} under moderate driving conditions can be well approximated by a linear parameter-varying model scheduled by the longitudinal speed. Experience of driving at low speed ($<3$m/s) suggest significant variation of the pneumatic trail $n_{sa}$, depending on the steering angle. 2.) The intended use of the model is control design for autonomous driving tasks in moderate driving conditions in urban environment. 3.) Experimental constraints concerning speed and space limits on the test field.
In order to satisfy space constraints the experiments are carried out in closed-loop, tracking a predefined trajectory. The region of operation is defined in terms of speed and steering angle. The designed trajectories systematically cover this two dimensional space. An example path is shown in Fig. \ref{fig:ZalaZoneGPSident}. The path is followed multiple times at different (approximately) constant speed between 1 m/s and 8 m/s and the signals are sampled with 50 Hz. The applied baseline path tracking control system is described in \cite{IFAC2020Nissan}. To ensure sufficient excitation for identification, the control signal is superimposed by a \emph{pseudo random binary signal} (PRBS) whose bandwidth (10 rad/s) exceeds the bandwidth of the lateral control system in normal driving conditions and whose amplitude is still bearable for the test pilots. In this way, 21 data sets $\{\mathcal{D}_i\}_{i=1}^{21}$ have been obtained.

\vspace{-2mm}
\section{Identification Results} \label{sec:ident}
\vspace{-3mm}
\subsection{Model structures} 
\vspace{-3mm}
Based on the overall description of the to-be-identified steering dynamics in Section \ref{sec:sys} and the measurement conditions detailed in Section \ref{sec:mes}, the data generating system is assumed to have the discrete-time form
\begin{subequations} \label{eq:datagen_sys}
\begin{align}
    x_{k+1} &= f(x_k,u_k), \\
    y_k &= h(x_k) + e_k,
\end{align}
\end{subequations}
where $k\in\mathbb{Z}$ is the discrete time corresponding to the sampling $t=kT_\mathrm{s}$, $u_k\in\mathbb{R}^2$ is the input signal composed as $u(k)=[\begin{array}{cc}  u_\mathrm{s}(k) & v(k) \end{array}]^\top$, $y_k=r_k \in\mathbb{R}$ is the yaw rate, $x_k\in\mathbb{R}^{n_\mathrm{x}}$ is the state variable which contains sampled versions of $v_{x}$, $v_{y}$ and $\delta$ and states related to the unknown components of the system, but its exact order $n_\mathrm{x}$ and composition is unknown, $e_k$ is the measurement noise process that is assumed to be colored with finite variance. The functions $f:\mathbb{R}^{n_\mathrm{x}+2}\rightarrow \mathbb{R}^{n_\mathrm{x}}$ and $h:\mathbb{R}^{n_\mathrm{x}}\rightarrow \mathbb{R}$ are potentially non-smooth.
In order to capture \eqref{eq:datagen_sys}, we consider to use three model structures:\vspace{-2.5mm}

\subsubsection{LTI-SS model:}A \emph{linear-time-invariant} (LTI) unstructured \emph{state-space} (SS) model of the form
\begin{subequations} \label{eq:LTISS}
\begin{align}
    \hat{x}_{k+1} &=A_\theta \hat{x}_k + B_\theta u_k + K_\theta \epsilon_k, \\
    \hat{y}_k &= C_\theta \hat{x}_k + \epsilon_k,
\end{align}
\end{subequations}
where $\epsilon_k$ is the innovation noise and assumed to be i.i.d.~with $\epsilon_k\sim \mathcal{N}(0,\sigma_\mathrm{e}^2)$, $\sigma_\mathrm{e}>0$, and $A_\theta,\ldots, K_\theta$ are matrices with appropriate dimensions representing the parameters $\theta$ of the model, see \cite{Ljung:1999} for further details.  Due to the fact that the measurement data is not applicable for \emph{nonlinear distortion analysis} \citep{7470636}, the linear model structure is used to assess how nonlinear the underlying process is. Note that the overall dynamics are expected to have a dominant nonlinear behavior as it is shown in \cite{9139282}, hence the LTI model structure is only used for comparison purposes. \vspace{-5.5mm}

\subsubsection{NL-GP model:} A non-parametric \emph{nonlinear} (NL) \emph{Gaussian process} (GP) model of the form
\begin{equation}
\hat{y}_k=\hat{g}(y_{k-1},\ldots,y_{k-n_\mathrm{a}},u_{k-1},\ldots,u_{k-n_\mathrm{b}})+\epsilon_k
\end{equation}
where $\hat{g}$ is the \emph{maximum a posteriori} function estimate under a GP prior $g\sim GP(0,\mathcal{K}_\theta)$, see \cite{Rasmussen2006} for further details. Here $\mathcal{K}_\theta$ is a kernel function parameterized in terms of its hyper-parameters $\theta$ which in fact represent the to-be-estimated parameters of the model. $\mathcal{K}_\theta$ also defines the function space in which the unknown dynamic relation of \eqref{eq:datagen_sys}, characterized by $\hat{g}$ in an input-output form, is estimated. $GP$ estimators are well-known for their efficiency to capture difficult nonlinear relationships from data \citep{Rasmussen2006}.

\subsubsection{NL-ANN model:} A nonlinear \emph{artificial neural network} (ANN) based SS model: 
\begin{subequations}
\label{eq:ANN}
\begin{align}
    \hat{x}_{k+1} &= f_\theta(\hat{x}_k,u_k), \\
    \hat{y}_k &= h_\theta(\hat{x}_k) + \epsilon_k,
\end{align}
\end{subequations}
with $\hat{x}_k \in \mathds{R}^{n_\mathrm{x}}$ the internal model state, $\epsilon$ is a colored noise process, $f_\theta$ and $h_\theta$ constructed and parameterized as a multi layered ANN with linear bypass and weight parameters $\theta$. The functions $f_\theta$ and $h_\theta$ are represented as ANNs to capute the expected complex dynamics.

\begin{figure}[t]
    \centering \vspace{-2mm}
    \includegraphics[width=0.7\linewidth]{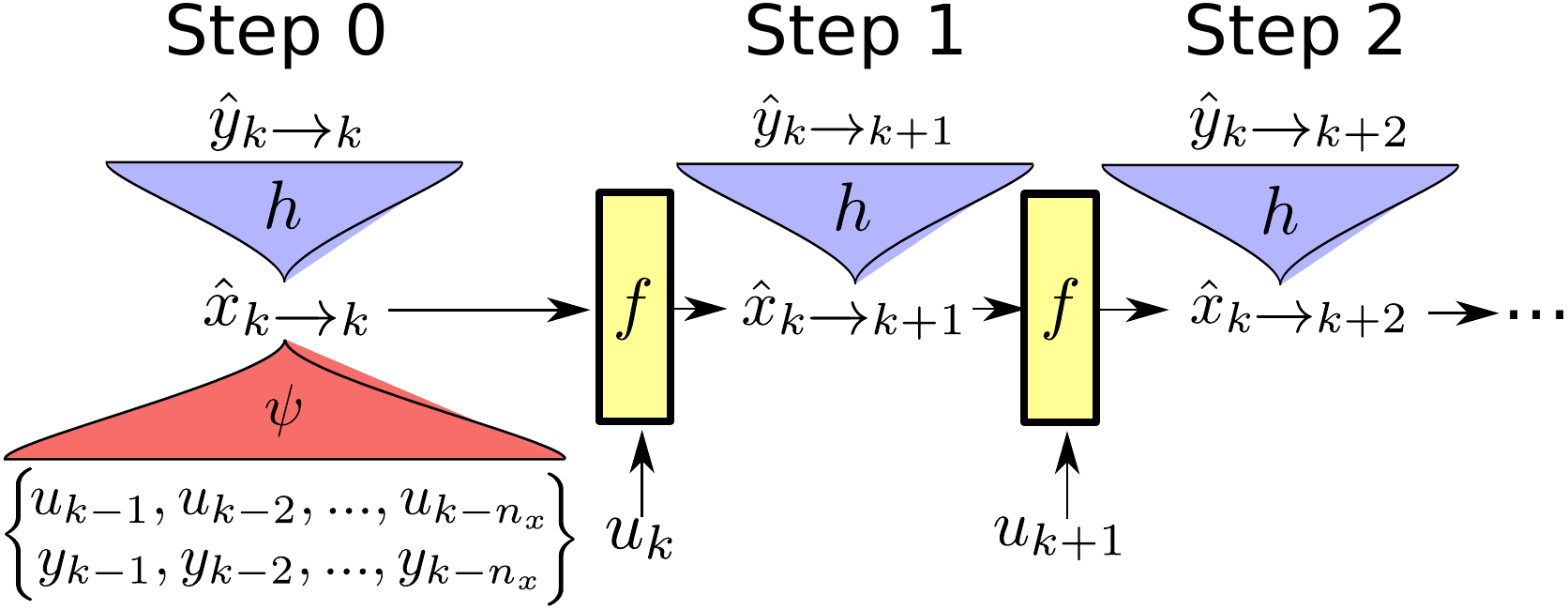}\vspace{-2mm}
    \caption{Estimation structure of the state-space encoder where the initial state $\hat{x}_{k\rightarrow k}$ is estimated by an encoder function $\psi$ based on past IO samples.} \vspace{-0mm}
    \label{fig:encoder}
\end{figure}

\vspace{-3mm}
\subsection{Model estimation}
\vspace{-2mm}
The LTI-SS model structure \eqref{eq:LTISS} is identified with state-space \emph{prediction error minimization} (PEM) \citep{Ljung:1999}  initialized using SS-ARX based subspace identification via the Matlab script {\tt ssest}. The effective model order was determined to be 10. Note that the used innovation noise model is rather general and it is adequate in the LTI case to provide unbiased estimates under colored output noise.

Due to the fact that the GP model assumes an \emph{nonlinear auto-regressive} noise structure (NARX), while the expected noise in the system has a colored OE form, the standard MAP estimate based on the 1-step ahead prediction error results in a seriously biased estimate, which in simulation performs worse than the estimated LTI model on the test data. To avoid the resulting bias, the GP estimator was trained on the $n$-step ahead prediction error, which in theory converges to the simulation error when $N\rightarrow \infty$. The kernel function was chosen to be a \emph{squared exponential}, and the hyper-parameters together with the selection of the orders $n_\mathrm{a}$ and $n_\mathrm{b}$ were tuned via cross validation using an $\ell_2$ loss function.

For the estimation of the NL-ANN-SS model \eqref{eq:ANN} a recently developed subspace-encoder estimation approach \cite{gerben2021a} is utilized. The main idea behind of the method is depicted in Fig. \eqref{fig:encoder}, where a state-sequence $x_k$  is estimated as $\hat{x}_{k \rightarrow k}$ using a nonlinear reconstructability map $\psi_\theta$ from past inputs and outputs that is parameterized as a deep neural network. Then, the state sequence is forward propagated in time through an ANN parametrization based $f_\theta$ and $h_\theta$ which results in a nonlinear (observability) mapping of  $x_{k \rightarrow k}$ to the future sequence of the output $\hat y_k,\ldots,\hat y_{k+n}$, which are denoted as $\{\hat y_{k \rightarrow  k + i}\}_{i=1}^n$ to emphasize that they are the predicted outputs based on the estimated state at time moment $k$. In fact, the estimation structure corresponds to  
\begin{subequations}
\begin{align}    
     \hat{y}_{k \rightarrow k + i} &:= h_\theta( \hat{x}_{k \rightarrow k +i}),\\
     \hat{x}_{k \rightarrow k + i+1} &:= f_\theta(\hat{x}_{k \rightarrow k + i},u_{k + i}),\\
     \hat{x}_{k \rightarrow k} &:= \phi_\theta(y_{k-n_\mathrm{x}:k-1},u_{k-n_\mathrm{x}:k-1}),
     \label{eq:encoder}
\end{align}
\end{subequations}
where $\hat x_{k \rightarrow  k + i}$ indicates $i$ recursive uses of $f_\theta$ to calculate
\begin{gather*}
    \hat{x}_{k \rightarrow k + i} = f_\theta(f_\theta( ... f_\theta(\hat{x}_{k \rightarrow k}, u_{k}),..., u_{k + i-2}), u_{k + i-1}).
\end{gather*}
By simultaneously optimizing $f_\theta,h_\theta, \psi_\theta$ based on the $\ell_2$ loss of the observability map based prediction 
\begin{equation}
    V_{\text{enc}}(\theta) \!=\! \frac{1}{2 N (n+1)} \sum_{k=1}^N \sum_{i=\tau_0}^{n+\tau_0} \|y_{k+i}-\hat{y}_{k \rightarrow k +i }  \|^2_2,
\end{equation}
with $\tau_0\geq0$ \emph{burn in} parameter, the estimator can be seen as a nonlinear extension linear subspace identification \cite{gerben2021a} or as a generalization of the multiple shooting method~\citep{ribeiro2019multiple-shooting}. Furthermore, due to the independence of the loss function on each section, the proposed method allows for \textit{i)} computational speedup by utilizing modern parallelization methods and \textit{ii)} the utilization of batch optimization methods (e.g. Adam~\citep{kingma2014adam}).
For the given identification problem, $\psi_\theta$, $f_\theta$ and $h_\theta$ are chosen as a two hidden layer neural networks with $64$ nodes per layer, $\tanh$ activation and a linear bypass. The parameters are initialized by sampling the uniform distribution $\mathcal{U}(\pm{1/\sqrt{n_{\text{in}}}})$ based on the number of inputs $n_{\text{in}}$ to the layer. 
For training, the following hyper-parameters are used: $n_\mathrm{x} = 40$, $\tau_0 = 0$, $n = 100$ with Adam batch optimization \citep{kingma2014adam} using learning rate of $\alpha = 10^{-3}$ and a batch size of $512$. Note that the hyper-parameters were manually optimized including the network depth and state-order. The most critical parameters to tweak where the batch size and the downsample factor (see next section) whereas the precise structure of the ANN had less prominent role. The detailed results for various choices of these parameters are not reported here due to space restrictions.

\vspace{-2mm}
\subsection{Results}
\begin{figure*}[t]\vspace{-3mm}
    \centering
    \subfigure[\tiny LTI-OE and LTI-SS on $\mathcal{D}_{14}$, $\text{NRMS}=50\%, 77\%$]{\includegraphics[width=5.95cm]{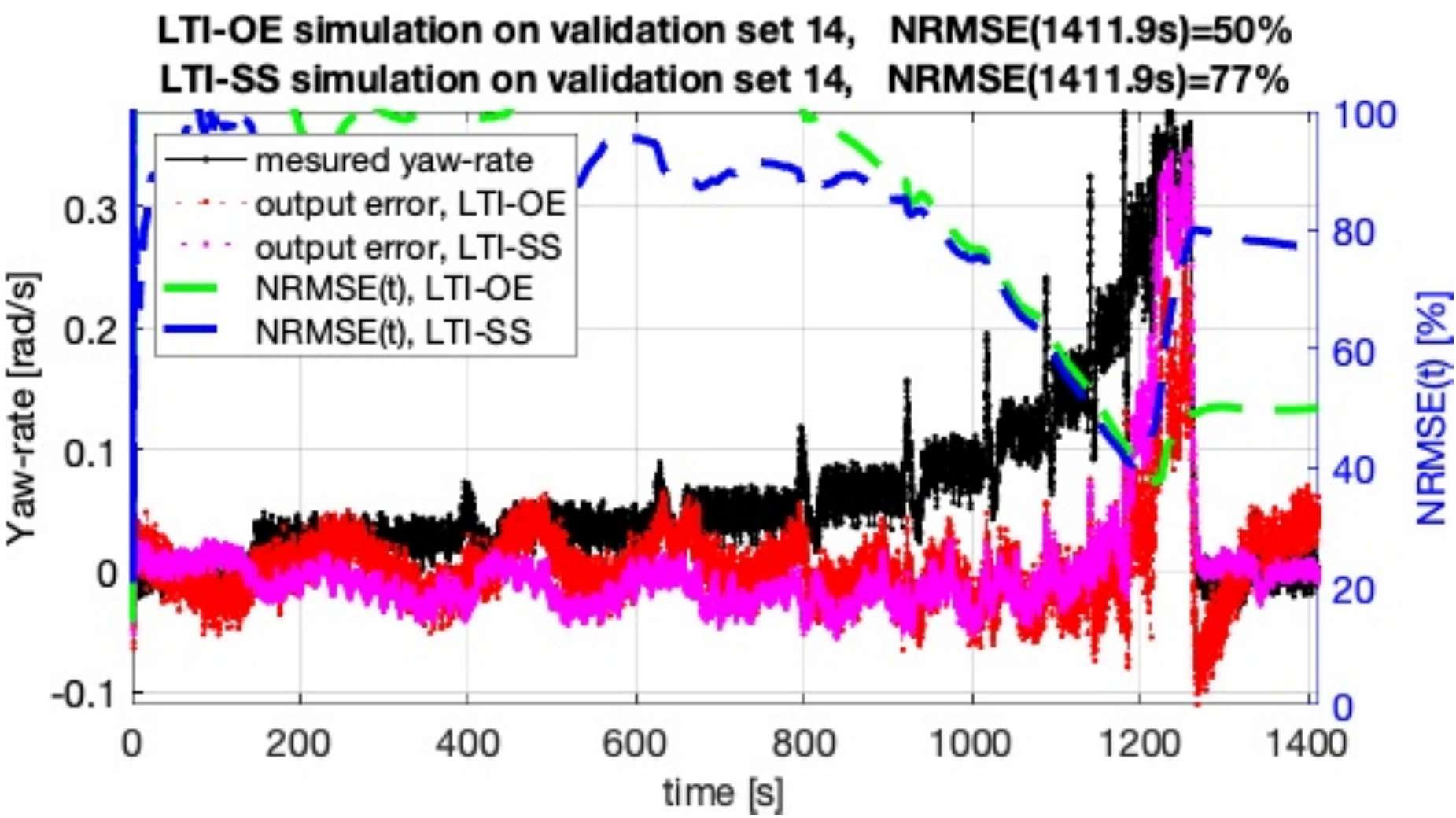}} 
    \subfigure[\tiny GP-OE on $\mathcal{D}_{14}$, $\text{NRMS}=82\%$]{\includegraphics[width=5.95cm]{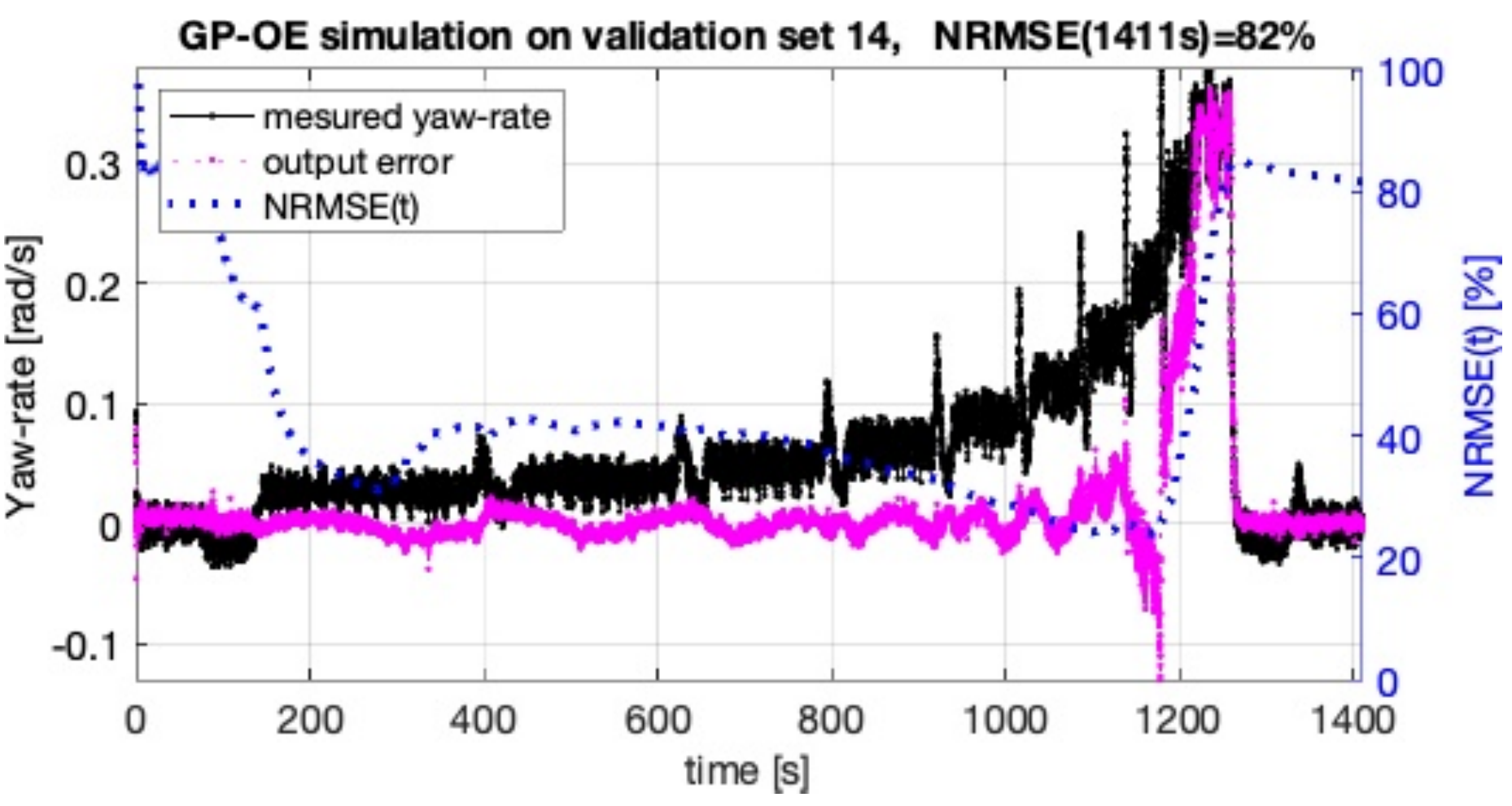}} 
    \subfigure[\tiny NL-ANN on $\mathcal{D}_{14}$, $\text{NRMS}=17\%$]{\includegraphics[width=5.95cm]{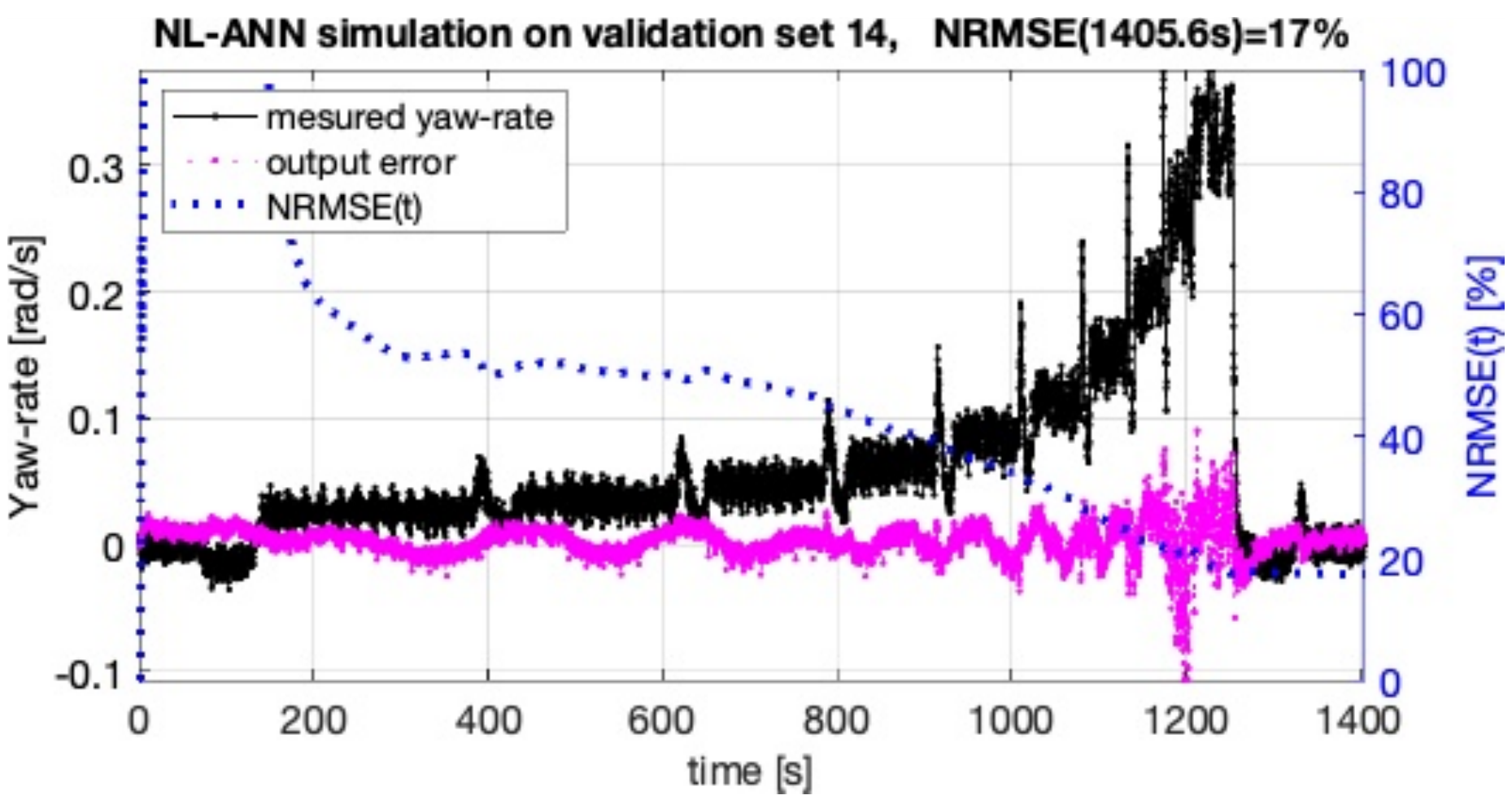}}\\
    \subfigure[\tiny LTI-OE and LTI-SS on $\mathcal{D}_{18}$, $\text{NRMS}=26\%, 21\%$]{\includegraphics[width=5.95cm]{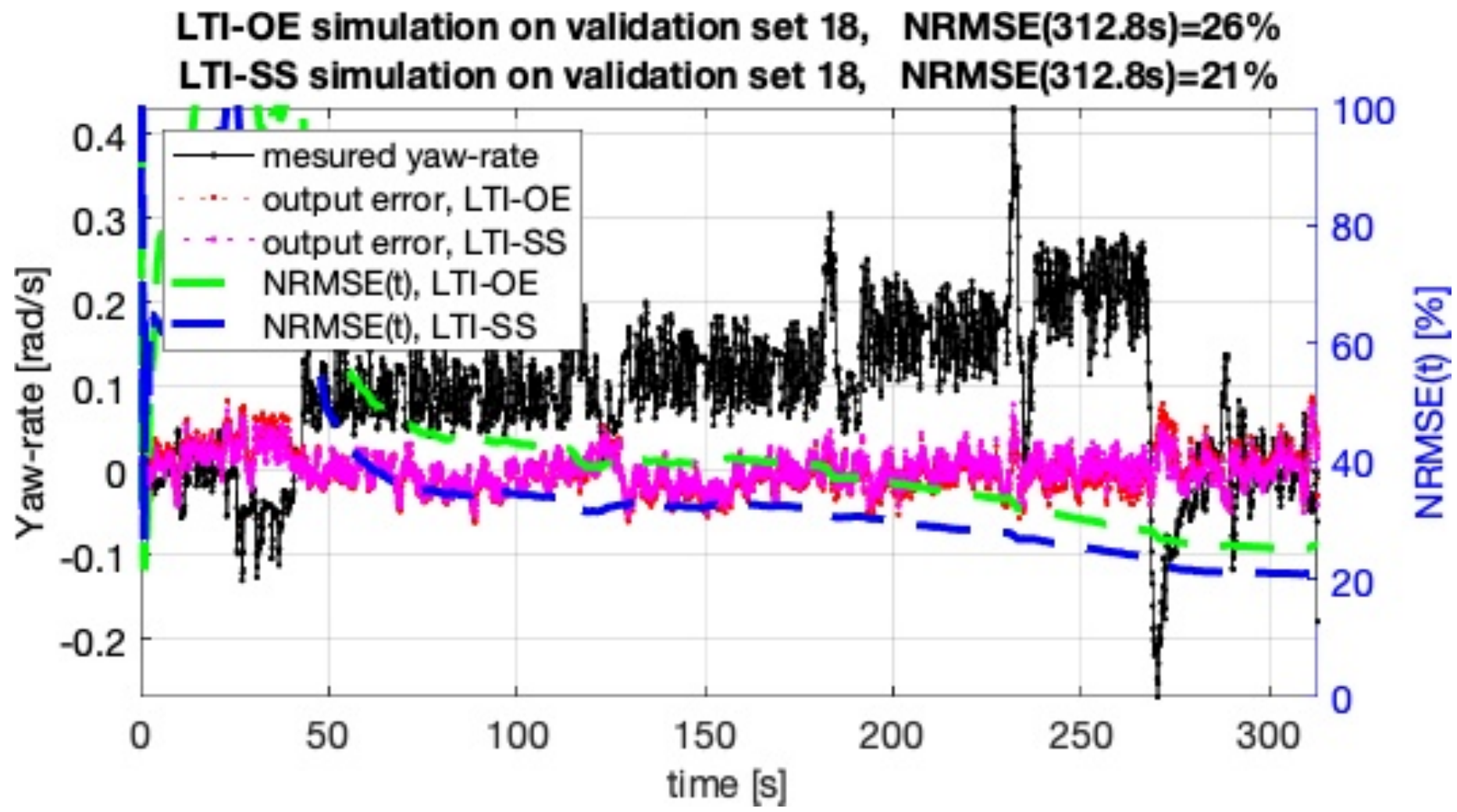}} 
    \subfigure[\tiny GP-OE on $\mathcal{D}_{18}$, $\text{NRMS}=18\%$]{\includegraphics[width=5.95cm]{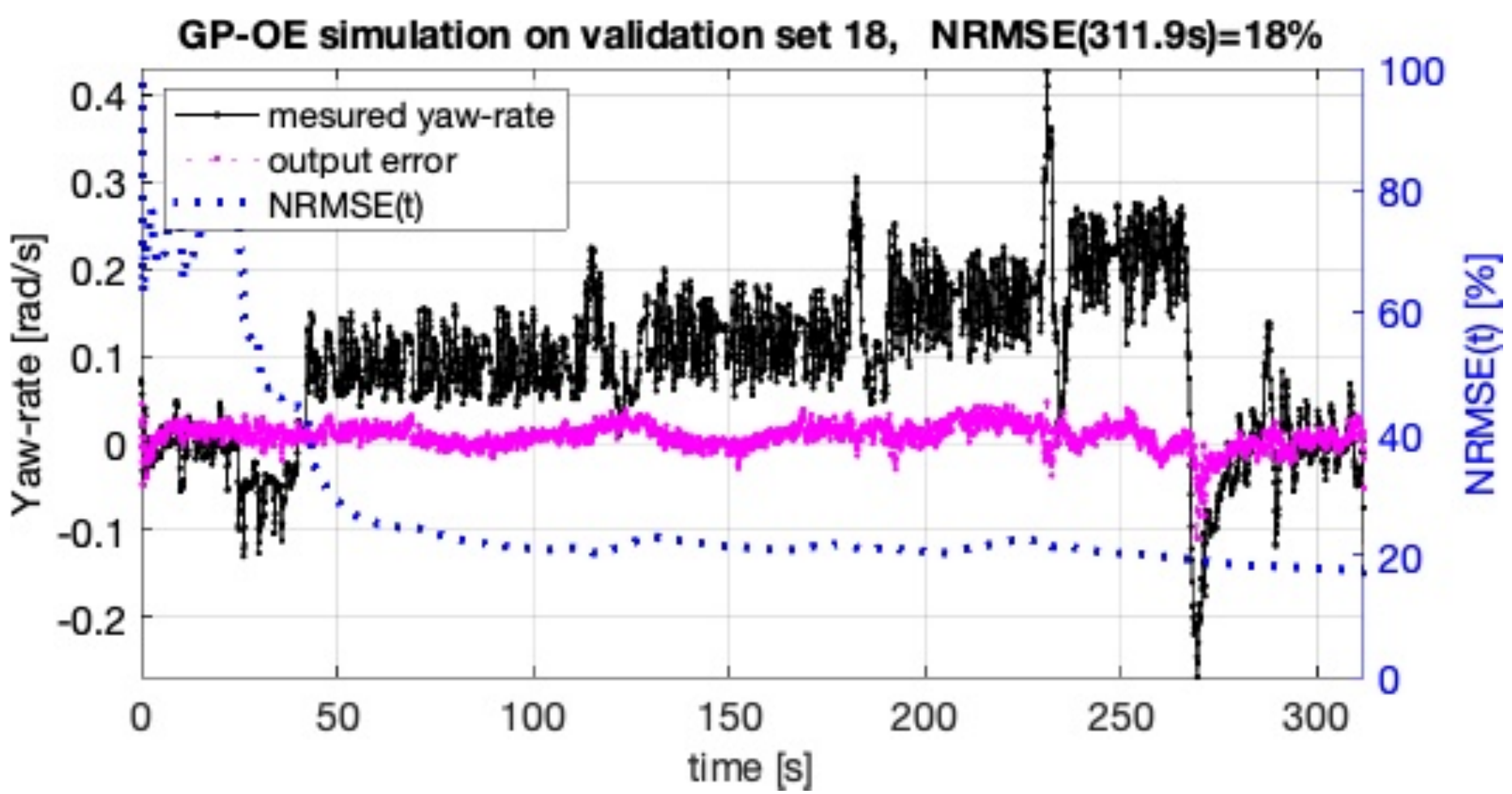}} 
    \subfigure[\tiny NL-ANN on $\mathcal{D}_{18}$, $\text{NRMS}=16\%$]{\includegraphics[width=5.95cm]{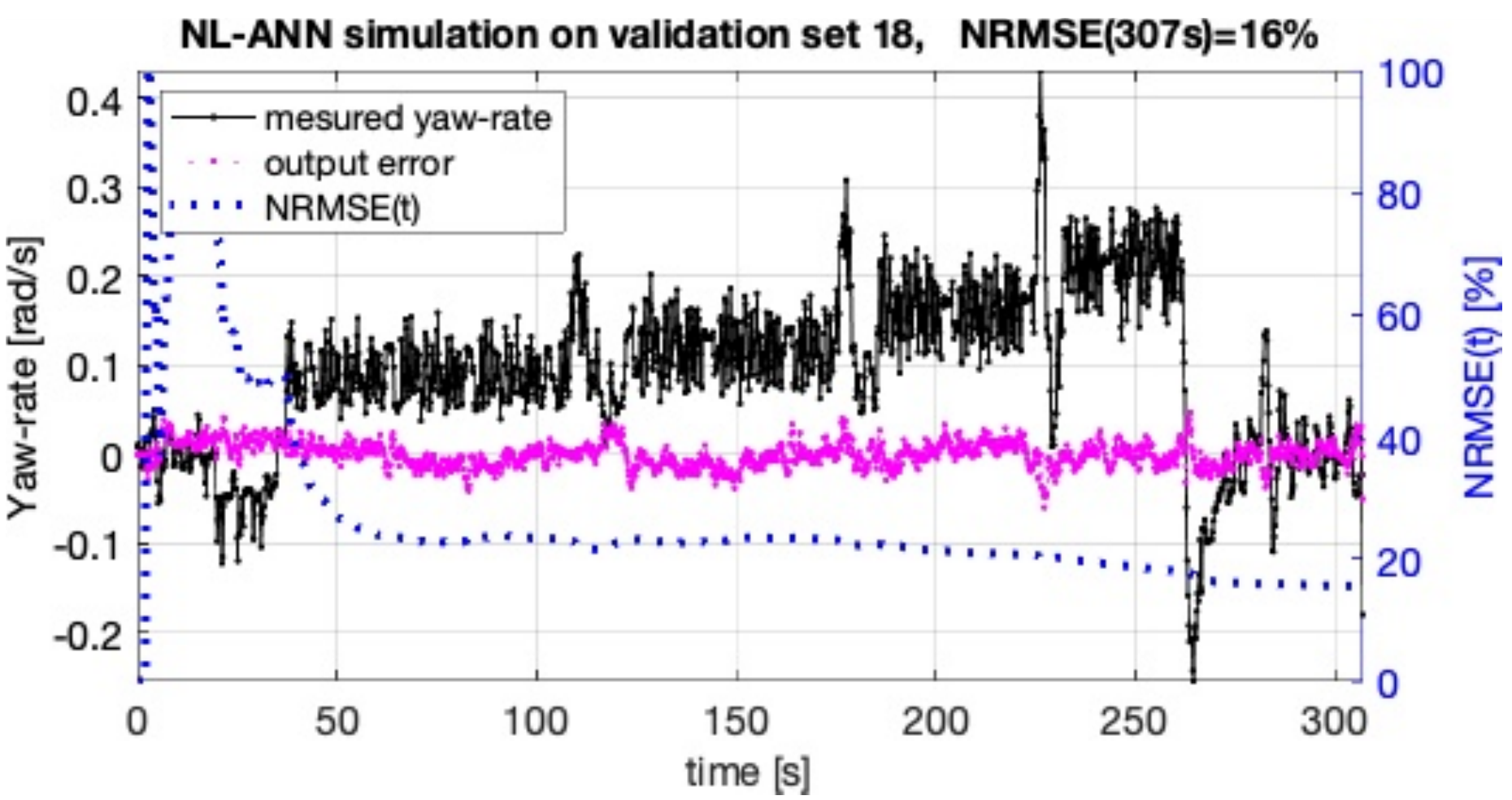}} 
    \vspace{-2mm}   
    \caption{Measured yaw-rate, error of the simulated model responses and evolution of the normalized RMSE on the test data. The models are compared at low speed $\sim$5 km/h (top row) and moderate speed $\sim$23 km/h (bottom row). }
    \label{fig:res}\vspace{-2mm}
\end{figure*}

\vspace{-2mm}
The results of the estimation based on the three approaches are given in Table \ref{table:results} and Figure \ref{fig:res}. The data has been subsampled with an FIR based anti-aliasing filter to sampling time $T_\mathrm{s}=0.1$ sec which results in negligible loss of the represented frequency content. For the NL-ANN-SS model, a subsampling to $T_\mathrm{s}=0.2$ sec was used to decrease the computational load across the 21 data sets and sets $\{\mathcal{D}_i\}_{i\in\{1, 3, 4, 7, 8, 9, 10, 11, 12, 15, 16, 18, 20\}}$ were used for training,  $\{\mathcal{D}_i\}_{i\in\{2, 6, 19\}}$ were used as validation sets to train the hyper-parameters and  $\{\mathcal{D}_i\}_{i\in\{5, 13, 14, 17\}}$ were used as test sets for evaluating the identified model in terms of the \emph{normalized root mean square error} (NRMSE) of the simulated model response reported in Table \ref{table:results}. The cumulative NRMSE up to sample $k\in\{1,\ldots,N\}$ is 
\begin{equation}
	\text{NRMSE}(k) = \frac{\sqrt{\sum_{j=1}^k \|y_j-\hat{y}_j\|^2_2}}{\sqrt{\sum_{j=1}^k \|y_j- \bar{y}\|^2_2}}
\end{equation}
where $\bar{y}$ is the mean of the signal $y$ and $N$ is the length of the data set. This allows in Figure \ref{fig:res} to show also the evolution of NRMSE$(k)$. 

The estimation of the LTI-SS model has been accomplished on $\{\mathcal{D}_2\}$ and the resulting model was tested on $\{\mathcal{D}_{14}\}$ as these sets represent responses for similar speed profile. As pure LTI estimation failed to provide reasonable results, Hammerstein-Wiener identification based analysis has shown that $u_\mathrm{s}$ is effected by a dead-zone nonlinearity in the region $[-0.13,0.17]$. Hence the LTI estimate was recomputed by first applying this dead-zone nonlinearity on $u_\mathrm{s}$ and the corresponding results are reported with LTI-SS$^\ast$. As it was noticed that {\tt ssets} has experienced numerical problems, the estimation was repeated using an LTI-OE model structure with $n_\mathrm{b}=[9,8]$, $n_\mathrm{f}=[6,5]$ and $n_\mathrm{k}=[1,0]$ (determined via residual analysis), which in terms of NRMSE has obtained better results. However, careful analysis of the simulation response for $\{\mathcal{D}_{14}\}$ in Figure \ref{fig:res} reveals that qualitatively neither of the models provide good approximation of the response due to serious oscillations of the OE model in the low yaw rate region and complete miss of the high-yaw-rate response. LTI models were also trained and respectively tested with higher speed (23 km/h) experiment data $\{\mathcal{D}_{11}\}$ and tested on $\{\mathcal{D}_{18}\}$, where the range of steering angle is smaller, hence the self-aligning torque is positive and keeps the system in the stable region. As a result, LTI models perform for such data sets significantly better. In overall, LTI identification shows serious inconsistency over the estimation data sets, with even poor results on validation sets with similar speed profile. This clearly shows the highly nonlinear and volatile nature of the involved dynamics for large steering angles.

NL-GP-OE model with $n_\mathrm{a}=n_\mathrm{b}=9$ was trained on $\{\mathcal{D}_{2}\}$ and tested on $\{\mathcal{D}_{14}\}$ for the low speed range (5 km/h), and trained on $\{\mathcal{D}_{17,19}\}$ and tested on $\{\mathcal{D}_{18}\}$ for the higher speed range (23 km/h). Due to the large amount of data, only selected batches were used for hyper-parameter tuning and training. The evolution of NRMSE in Figure \ref{fig:res} shows that the GP model performs similar to NL-ANN except for the sensitive range of the system, i.e., at low speed and sharp cornering, where it completely misses the high-yaw-rate response. One drawback of the GP model compared to NL-ANN that its state-space $[\hat y_{k-1}\ \cdots\ u^\top_{k-n_b}]^\top$ is fixed in advance, while the NL-ANN have a large model order, can work with much higher amount of data and can optimize its state vector to have a more parsimonious representation of the past. This allows the state-space encoder method to obtain superior results with relatively negligible simulation error on the test data.
Note that other nonlinear black-box identification methods such as \citep{Toth20IFACa} have been also applied on the data set, but the obtained results ($75.78\%$ on $\mathcal{D}_{14}$ and  $20.51\%$ on $\mathcal{D}_{18}$, but $149.14\%$ on $\mathcal{D}_{21}$) have not been competitive with the presented approaches.

\begin{table}[t]
\caption{(in)Validation results of the models.}  \label{table:results} \vspace{-1mm}
\begin{center}
\small{
\begin{tabular}{ccccc}
\hline
   Measure/Data  &           LTI-SS$^\ast$ & LTI-OE & NL-GP  & NL-ANN  \\
     \hline
   NRMSE(N) $\mathcal{D}_{14}$ & 77\% & 50\% & 82\% & 17\%\\
   NRMSE(N) $\mathcal{D}_{18}$ & 21\% & 26\% & 18\% & 16\%\\
\hline
    \end{tabular}
  }
\end{center}
\end{table}

\vspace{-2mm}
\section{Conclusions}
\vspace{-2mm}
In this paper, we have discussed the identification of the steering system of a Nissan Leaf based autonomous car. Based on the provided description of the system and the obtained experimental data, we have shown that the underlying dynamics are highly nonlinear and hence difficult to be captured by linear models even under constant lateral speed. To model the challenging behavior of the system, a Gaussian process (GP) based non-parametric estimator and a subspace encoder based neural network approach have been compared, where based on the invalidation results, only the latter approach could successfully capture the dynamics, due to the computational efficiency of the encoder even under high model order and the more parsimonious form of the ANN-SS model structure. Future work involves the physical interpretation of the obtained model estimates, further refinement of the involved model structure choices in terms of kernels, order selection and sparsification, and inclusion of parameterized noise models with the final objective of  using the estimated model in designing a model predictive control based autopilot. \vspace{-2mm}

\bibliography{sysid21_Nissan_ver_tr12}             

\end{document}